\input iopppt.modifie
\input xref
\input epsf


\jnlstyle

\jl{1}

\letter{Conformal off-diagonal boundary density profiles on a
semi-infinite strip}[Letter to the Editor]

\author{L~Turban}[Letter to the Editor]

\address{Laboratoire de Physique des Mat\'eriaux, Universit\'e Henri Poincar\'e
(Nancy~I), BP~239,  F--54506~Vand\oe uvre l\`es Nancy Cedex, France}


\abs
The off-diagonal profile $\phi_{\rm od}^{\rm b}(v)$ associated with a local 
operator $\hat{\phi}(v)$ (order parameter or energy density) close to the 
boundary of a semi-infinite strip with width $L$ is obtained at criticality 
using conformal methods. It involves the surface exponent $x_\phi^{\rm s}$ and 
displays a simple universal behaviour which crosses over from surface
finite-size  scaling when $v/L$ is held constant to corner finite-size scaling
when $v/L\to 0$.  \endabs

\vglue1cm

\pacs{0550, 6842}

\submitted

\date

The finite-size behaviour of order parameter or energy density profiles has
been the subject of much interest during the last two decades following the
work of Fisher and de Gennes~\cite{fisher78}. These profiles have been
studied in the vivinity of the critical point in the mean-field
approximation~\cite{binder83}, using field-theoretical
methods~\cite{diehl86} and through exact
solutions~\cite{auyang80,bariev88}. Such profiles display universal behaviour
at criticality and in two-dimensional systems they can be deduced from
ordinary scaling and covariance under conformal
transformation~[6--17]. A
short review can be found in reference~\cite{henkel99}.

With symmetry-breaking boundary conditions, one may consider diagonal order
parameter profiles~\cite{burkhardt85}, i.e., ground-state
expectation values. Otherwise, off-diagonal profiles can be used with any
type of boundary  conditions~\cite{turban97}. 
For the order parameter with Dirichlet boundary
conditions, off-diagonal matrix elements must be considered since a diagonal
order parameter profile then vanishes for symmetry reasons.

On a strip with fixed boundary conditions at $v=0$ and $v=L$ the
diagonal order-parameter profile $\phi(v)$ associated with an operator
$\hat{\phi}$ takes the following form at criticality~\cite{burkhardt85}
$$
\phi(v)=\langle0\vert\hat{\phi}(v)\vert0\rangle={\cal
A}\left[{L\over\pi}\sin\left({\pi v\over L}\right)\right]^{-x_\phi}\qquad
0<v<L\,. 
\label{e1}
$$
The exponent $x_\phi$ is the bulk scaling dimension of $\hat{\phi}$,
$\vert0\rangle$ is the ground state of the Hamiltonian ${\cal H}=-\ln{\cal T}$
where ${\cal T}$ denotes the row-to-row transfer operator on the strip. 
When $L\to\infty$ with a fixed value of the ratio $v/L$, one obtains
the bulk finite-size scaling behaviour $\phi(L)\sim L^{-x_\phi}$. When
$L\to\infty$ while keeping $v$ fixed, one obtains the profile
$\phi(v)\sim v^{-x_\phi}$ on the half-plane with fixed boundary conditions which
is a consequence of ordinary scaling. Actually the profile on the strip
in~\ref{e1}  follows from the profile on the half-plane through the logarithmic
conformal transformation $w=(L/\pi)\ln z$~\cite{burkhardt85}.

The off-diagional critical profile with general symmetric boundary conditions at
$v=0$ and $L$ is obtained as~\cite{turban97} 
$$
\phi_{\rm od}(v)=\langle\phi\vert\hat{\phi}(v)\vert 0\rangle
\sim\left({L\over\pi}\right)^{-x_\phi}
\left[\sin\left({\pi v\over L}\right)\right]^{x_\phi^{\rm s}-x_\phi} 
\qquad 0<v<L\,, 
\label{e2}
$$
where $\vert\phi\rangle$ is the lowest excited state of $\cal H$
leading to a non-vanishing matrix element. Besides the bulk exponent $x_\phi$
the off-diagonal profile involves the surface scaling dimension $x_\phi^{\rm s}$
of the operator $\hat{\phi}$. It can be identified by considering the
transformation of the connected two-point correlation function ${\cal
G}_{\phi\phi}^{\rm con}(z_1,z_2)$ from the half-plane to the strip under the
logarithmic conformal mapping. For the order parameter with fixed boundary
conditions, $x_\phi^{\rm s}=0$, and~\ref{e2} gives an off-diagonal profile in
agreement with~\ref{e1}.  When $L\to\infty$, equation~\ref{e2} shows the
crossover from bulk finite-size scaling $\phi_{\rm od}(L)\sim
L^{-x_\phi}$ when the ratio $v/L$ is constant, to surface finite-size scaling
$\phi_{\rm od}(L)\sim L^{-x_\phi^{\rm s}}$ when $v$ is constant, i.e.,
when $v/L\to0$.

{\par\begingroup\medskip
\epsfxsize=10truecm
\topinsert
\centerline{\epsfbox{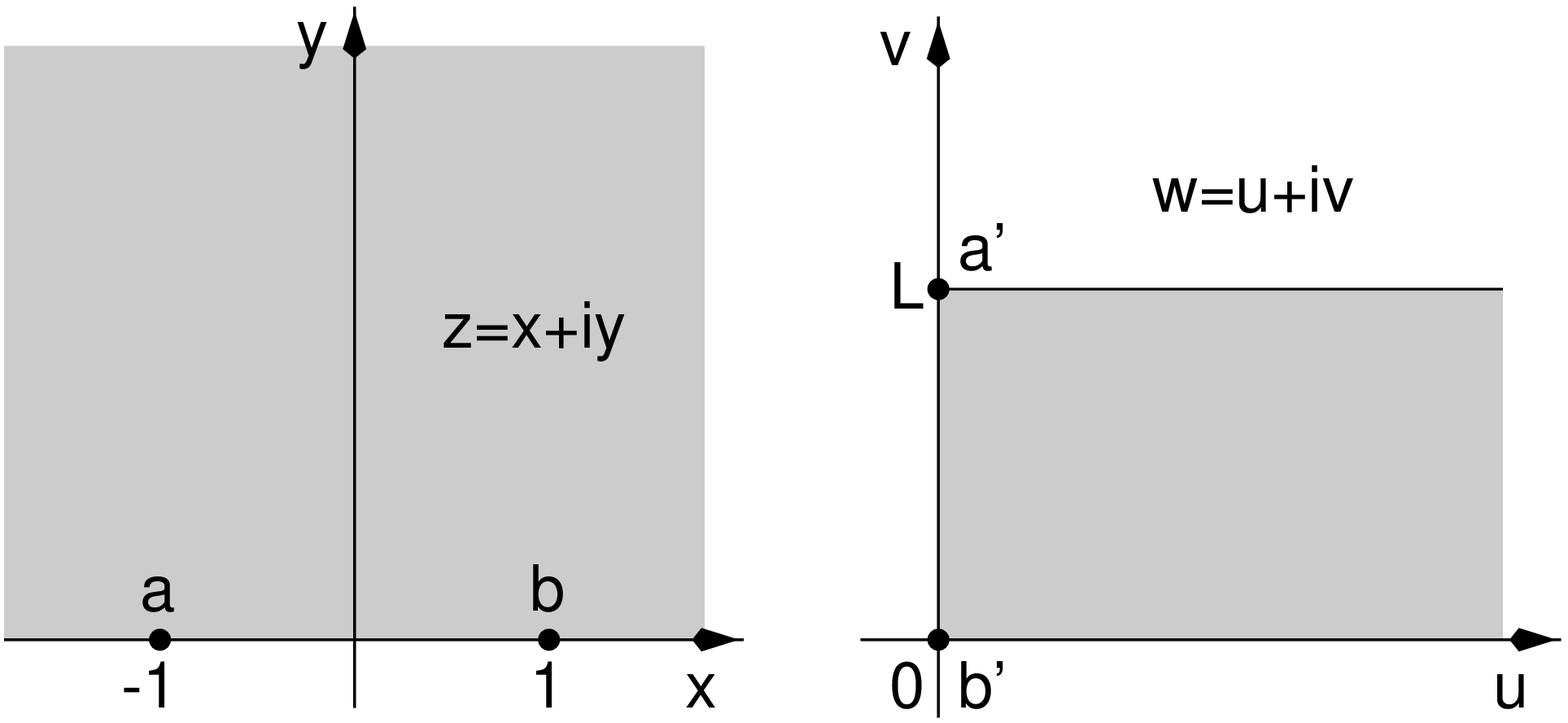}}
\vglue5mm
\figure{Conformal mapping of the half-plane $y>0$ on the semi-infinite strip
$u>0$, $0<v<L$. } 
\endinsert
\endgroup
\par}

Let us now consider a half-strip in the $(u,v)$-plane with $0<u<\infty$,
$0<v<L$ and uniform boundary conditions. If one crosses the semi-infinite strip
at $u\gg L$ the behaviour of the off-diagonal profile will be the same as for the
infinite strip in equation~\ref{e2}. A different behaviour is expected close to
the boundary of the semi-infinite strip at $u\ll L$. The profile should then
involve the surface exponent $x_\phi^{\rm s}$ and, when $v\ll L$ or $L-v\ll L$,
the corner exponent $x_\phi^{\rm c}$. 

In order to obtain the profiles on the semi-infinite strip, we
consider the transformation of the connected two-point correlation
function ${\cal G}_{\phi\phi}^{\rm con}(z_1,z_2)$ from the half-plane
$z=x+\i y$, $y>0$ to the half-strip $w=u+\i v$, $0<u<\infty$, $0<v<L$ 
as shown in figure~1. The two geometries are related by the conformal 
transformation~\cite{burkhardt85}
$$
z=\cosh\left({\pi w\over L}\right)
\label{e3}
$$
or
$$
x=\cosh\left({\pi u\over L}\right)\cos\left({\pi v\over L}\right)\qquad
y=\sinh\left({\pi u\over L}\right)\sin\left({\pi v\over L}\right)\,.
\label{e4}
$$
Going from the half-plane to the half-strip, the dilatation factor is
given by: 
$$
b(z)=\left\vert{\d z\over\d w}\right\vert=\left\vert{\pi\over L}
\sinh\left({\pi w\over L}\right)\right\vert=
{\pi\over L}\left[\sinh^2\left({\pi u\over L}\right)
+\sin^2\left({\pi v\over L}\right)\right]^{1/2}\,.
\label{e5} 
$$

At criticality, the form of ${\cal G}_{\phi\phi}^{\rm con}(z_1,z_2)$ is
strongly constrained by conformal invariance. Using an infinitesimal special
conformal transformation which preserves the surface geometry, one obtains a
system of partial differential equation for the connected two-point 
correlation function on the half-plane, from which the following scaling 
form is deduced~\cite{cardy84a}: 
$$ 
\fl
{\cal G}_{\phi\phi}^{\rm con}(z_1,z_2)=(y_1y_2)^{-x_\phi}g(\omega)
\qquad \omega={(x_2-x_1)^2+(y_2-y_1)^2\over y_1y_2}\;. 
\label{e6}
$$
We are mainly interested in the behaviour of the profile close to the boundary of
the half-strip. Thus we may consider the correlations between two points, the
first close to the boundary located at $u=0$ and the second far from it:  
$$
{u_1\over L}\ll1\qquad 0<v_1<L\qquad {u_2\over L}\gg 1\qquad 0<v_2<L\,.
\label{e7}
$$
Considering \ref{e4} and \ref{e5} in the limits of equation~\ref{e7} we have: 
$$\eqalign{
&x_2-x_1\simeq{1\over2}\exp\left({\pi u_2\over L}\right)\cos\left({\pi
v_2\over L}\right)\quad 
y_2-y_1\simeq{1\over2}\exp\left({\pi u_2\over L}\right)\sin\left({\pi
v_2\over L}\right)\cr 
&y_1y_2\simeq{\pi u_1\over 2L}\sin\left({\pi v_1\over L}\right)
\sin\left({\pi v_2\over L}\right)\exp\left({\pi u_2\over L}\right)\cr 
&b(z_1)\simeq{\pi\over L}\sin\left({\pi v_1\over L}\right)\quad
b(z_2)\simeq{\pi\over L}\exp\left({\pi u_2\over L}\right)\cr}
\label{e8}
$$
Thus the crossover variable $\omega$ defined in~\ref{e6} takes the form
$$
\omega\simeq{L\exp(\pi u_2/L)\over2\pi u_1\sin(\pi v_1/L)\sin(\pi v_2/L)}\gg1\,.
\label{e9}
$$
In this limit, ordinary scaling leads to $g(\omega)\sim\omega^{-x_\phi^{\rm
s}}$ so that, in the half-plane geometry:  
$$ 
{\cal G}_{\phi\phi}^{\rm con}(z_1,z_2)\sim{(y_1y_2)^{x_\phi^{\rm s}-x_\phi}
\over[(x_2-x_1)^2+(y_2-y_1)^2]^{x_\phi^{\rm s}}}\,.  
\label{e10} 
$$
The conformal mapping~\ref{e3} leads to the correlation
function in the half-strip geometry~\cite{cardy84b}  
$$
{\cal G}_{\phi\phi}^{\rm con}(w_1,w_2)\sim b(z_1)^{x_\phi}b(z_2)^{x_\phi}
{\cal G}_{\phi\phi}^{\rm con}(z_1,z_2)\,.
\label{e11}
$$
Making use of~\ref{e8} in equations~\ref{e10} and~\ref{e11}, we
finally obtain:
$$\fl
{\cal G}_{\phi\phi}^{\rm con}(w_1,w_2)\!\sim\! u_1^{x_\phi^{\rm s}-x_\phi}
\!\left[\!\left({2\pi\over L}\right)\!
\sin\!\left({\pi v_1\over L}\right)\!\right]^{x_\phi^{\rm s}}\!\!
\left({L\over\pi}\right)^{\!\!-x_\phi}\!\!
\left[\sin\left({\pi v_2\over L}\right)\right]^{x_\phi^{\rm s}-x_\phi}
\!\!\!\!\!\!\!\!\!\exp\left(\!-{\pi x_\phi^{\rm s}u_2\over L}\!\right). 
\label{e12}
$$

In order to identify the different contributions to the two-point correlation
function in~\ref{e12}, we can rewrite it using the row-to-row transfer
operator ${\cal T}$ on the strip with width $L$. The two-point correlation
function on the semi-infinite strip reads
$$\fl
{\cal G}_{\phi\phi}(w_1,w_2)=
{\langle B\vert\hat{\phi}(v_1){\cal T}^{u_2}\hat{\phi}(v_2)\vert 0\rangle
\over\langle B\vert{\cal T}^{u_2}\vert 0\rangle} 
={\sum_n\langle B\vert\hat{\phi}(v_1)\vert n\rangle
\exp(-u_2E_n)\langle n\vert\hat{\phi}(v_2)\vert 0\rangle
\over\langle B\vert 0\rangle\exp(-u_2E_0)}
\label{e13}
$$
where $\vert 0\rangle$ is the ground state of ${\cal H}$ which is selected by
the transfer from $u=u_2$ to $u=\infty$ and $\vert B\rangle$ is a state vector
appropriate for the boundary conditions at $u=0$. In the case of free boundary
conditions, it describes the free summation over the boundary states. In the
last expression the summation is over the complete set of eigenstates $\vert
n\rangle$ of ${\cal H}$ with eigenvalues $E_n$.

The connected two-point correlation function is then obtained by substracting the
ground-state contribution to the eigenstate expansion:
$$\eqalign{
{\cal G}_{\phi\phi}^{\rm con}(w_1,w_2)&={\cal G}_{\phi\phi}(w_1,w_2)
-{\langle B\vert\hat{\phi}(v_1)\vert 0\rangle\over\langle B\vert 0\rangle}
\langle 0\vert\hat{\phi}(v_2)\vert 0\rangle\cr
&\simeq{\langle B\vert\hat{\phi}(v_1)\vert\phi\rangle\over\langle B\vert
0\rangle} \langle\phi\vert\hat{\phi}(v_2)\vert 0\rangle\exp[-u_2(E_\phi-E_0)]
\cr}
\label{e14}
$$ 
In the last expression we took into account the condition $u_2\gg L$. In this
limit, the eigenstate expansion is dominated by the contribution of the
lowest excited state $\vert\phi\rangle$ of ${\cal H}$ for which the matrix
elements are non-vanishing. 

Comparing with the conformal expression in \ref{e12}, we can read the
gap-exponent relation~\cite{cardy84b} in the exponential factor, $E_\phi-E_0=\pi
x_\phi^{\rm s}/L$, and the off-diagonal profile at $u_2\gg L$ in
agreement with equation~\ref{e2}. The remaining part can be then identified as
the boundary profile on the half-strip at $u_1\ll L$ and we obtain  
$$
\phi_{\rm od}^{\rm b}(v)=
{\langle B\vert\hat{\phi}(v)\vert\phi\rangle\over\langle B\vert 0\rangle}
\sim\left[{\pi\over L}\sin\left({\pi v\over L}\right)\right]^{x_\phi^{\rm s}}
\qquad 0<v<L\,.
\label{e15}
$$

In the case of the two-dimensional Ising model with free boundary conditions
${\cal H}$ is the Hamiltonian of the Ising model in a transverse field; if one
associates the Pauli spin operator $\sigma_l^x$  $(l=1,L)$ with the
order parameter, then the boundary state vector $\vert B\rangle$ is
explicitly given by~\cite{barber84}  
$$
\vert B\rangle=\prod_{l=1,L}{1\over\sqrt{2}}(\vert\sigma_l^x=+1\rangle
+ \vert\sigma_l^x=-1\rangle)=\prod_{l=1,L}\vert\sigma_l^z=+1\rangle\,. 
\label{e16}
$$
Both $\vert 0\rangle$ and $\vert B\rangle$ are even under the operator
$P=\prod_{l=1,L}\sigma_l^z$~\cite{barber84}. 
In the expression of the order parameter profile the state
$\vert\phi\rangle=\vert\sigma\rangle$, which contains a single fermionic
excitation, is odd under $P$. At $l=1$, the order parameter profile coincides
with the corner magnetization obtained in reference~\cite{barber84}. The
surface magnetic exponent is $x_\sigma^{\rm s}=1/2$~\cite{mccoy73}.  
For the energy density profile the
state $\vert\phi\rangle=\vert\epsilon\rangle$ contains two fermionic
excitations and is even under $P$. The surface energy exponent is then
$x_\epsilon^{\rm s}=2$~\cite{gehlen86,cardy86}.

When $L\to\infty$ with a fixed $v/L$ value, one obtains the surface
finite-size scaling behaviour $\phi_{\rm od}^{\rm b}(L)\sim L^{-x_\phi^{\rm
s}}$ while keeping $v$ constant leads to the corner finite-size scaling
behaviour 
$$
\phi_{\rm od}^{\rm b}(L)\sim v^{x_\phi^{\rm s}}L^{-2x_\phi^{\rm s}}\qquad
v\ll L\,.
\label{e17}
$$
Thus the corner exponent $x_\phi^{\rm c}(\pi/2)$ is given by $2x_\phi^{\rm
s}$. This result is in agreement with the general expression $x_\phi^{\rm
c}(\theta)=\pi x_\phi^{\rm s}/\theta$ for a corner with opening angle $\theta$,
which also follows from conformal invariance~\cite{cardy84a,barber84}.

\ack  The Laboratoire de Physique des Mat\'eriaux is Unit\'e
Mixte de Recherche CNRS No 7556.

\numreferences

\bibitem{fisher78}{Fisher M E and de Gennes P G 1978} {\it C. R. Acad. Sci. 
Paris} B {\bf 287} 207 

\bibitem{binder83}{Binder K 1983} {\it Phase Transitions and Critical
Phenomena} vol~8 ed C~Domb and J~L~Lebowitz (London: Academic Press) p~1

\bibitem{diehl86} {Diehl H W 1986} {\it Phase Transitions and Critical
Phenomena} vol~10 ed C~Domb and J~L~Lebowitz 
(London: Academic Press) p~75

\bibitem{auyang80}{Au-Yang H and Fisher M E 1980} {\PR} {\rm B} {\bf 21} 3956

\bibitem{bariev88}{Bariev R Z 1988} {\it Theor. Math. Phys.} {\bf 77} 1090
 
\bibitem{burkhardt85}{Burkhardt T W and Eisenriegler E 1985} {\JPA} {\bf 18}
L83 

\bibitem{burkhardt87a}{Burkhardt T W and Cardy J L 1987} {\JPA} {\bf 20}
 L233
 
\bibitem{burkhardt87b}{Burkhardt T W and Guim I 1987} {\PR} B
{\bf 36} 2080   

\bibitem{cardy90}{Cardy J L 1990} {\PRL} {\bf 65} 1443

\bibitem{burkhardt91a}{Burkhardt T W and Xue T 1991} {\PRL} {\bf 66} 895 

\bibitem{burkhardt91b}{Burkhardt T W and Xue T 1991} {\NP} B {\bf 354} 653

\bibitem{burkhardt94}{Burkhardt T W and Eisenriegler E 1994} {\NP} B {\bf 424}
487    

\bibitem{turban97}{Turban L and Igl\'oi F 1997} {\JPA} {\bf 30} L105

\bibitem{igloi97}{Igl\'oi F and Rieger H 1997} {\PRL} {\bf 78} 2473

\bibitem{karevski97}{Karevski D and Henkel M 1997} {\PR} B {\bf 55} 6429   

\bibitem{carlon98}{Carlon E and Igl\'oi F 1998} {\PR} B {\bf 57} 7877   

\bibitem{karevski00}{Karevski D, Turban L and Igl\'oi F 2000} {\JPA} {\bf 33}
2663

\bibitem{henkel99} {Henkel M 1999} {\it Conformal Invariance and Critical
Phenomena} (Berlin: Springer) p 346  

\bibitem{cardy84a}{Cardy J L 1984} {\NP} B {\bf 240} 514

\bibitem{cardy84b}{Cardy J L 1984} {\JPA} {\bf 17} L385

\bibitem{barber84}{Barber M N, Peschel I and Pearce PA 1984} {\it J. Stat.
Phys.} {\bf 37} 497 

\bibitem{mccoy73}{McCoy B M and Wu T T 1973} {\it The Two-Dimensional Ising
Model}\ (Cambridge: Harvard Uni\-ver\-si\-ty Press) p 132

\bibitem{gehlen86}{von Gehlen G and Rittenberg V 1986} {\JPA} {\bf 19} L631

\bibitem{cardy86}{Cardy J L 1986} {\NP} B {\bf 275} 200

\vfill\eject\bye